# Sand-Filler Structural Material with Low Content of Polyethylene Binder

Haozhe Yi,[1] Kiwon Oh,[2] Rui Kou,[1] Yu Qiao,[1,2,*]

[1] *Department of Structural Engineering, University of California – San Diego, La Jolla, CA 92093-0085, U.S.A.*

[2] *Program of Materials Science and Engineering, University of California – San Diego, La Jolla, CA 92093, U.S.A.*

[*] *Corresponding author. Phone: +1-858-534-3388; Email: yqiao@ucsd.edu*

**Abstract:** Currently, most of the waste plastics cannot be recycled, causing serious environmental concerns. In this research, we investigated a compaction formation technology to fabricate structural materials with thermoplastic binders. When the compaction pressure was 70~100 MPa, with only ~10 wt% polyethylene binder, the flexural strength was greater than that of typical steel-reinforced concrete, suitable to many construction applications. Because construction materials are tolerant to impurities, our work may provide a promising opportunity to recycle waste plastics and to reduce the portland cement production.

## 1. Introduction

In order to save energy and to reduce carbon emission, a major effort is to develop "green cement" – construction materials that can largely replace ordinary portland cement (OPC) [1]. Production of OPC demands calcination of mixed limestone, clay, and gypsum at ~1550 °C [2]. The mineral decomposition releases more than 800 kg carbon dioxide ($CO_2$) per ton of manufactured cement [3], and uses approximately 4 GJ energy [4]. Currently, in every ~15 parts of $CO_2$ generated from human-related activities, 1 part is from OPC manufacturing [5–9]. In every 10 parts of industrial energy consumption, ~1 part is associated with OPC processing and transportation [10]. Conventional "green cements", such as fly ash modified OPC [11] and alkali-activated geopolymers [12], are under extensive study. Another important "green cement" is polymer cement [13]. It is formed by mixing sand/soil particles and a polymer binder. This process emits almost no $CO_2$ and consumes little energy. However, conventional mixing relies on shear motion to distribute the binder phase, for which the binder amount needs to be larger than ~15 wt% [14,15]. A lower binder content usually results in a much reduced structural strength. Because the polymer binder is relatively expensive ($1-3 per kg) [16], traditional polymer cements cannot compete against OPC on cost. The applications of polymer cement are limited to small-sized high-end markets, such as waste chemical storage, water pipes, floor tiles, etc. [17–21], mainly because of their high strength, high corrosion resistance, and aesthetic features.

Recently, we performed investigation on the compaction self-assembly (CSA) technology to manufacture particulate composites with reduced binder contents [22–25]. In CSA, no long-range motion of materials components is involved. Premixed sand/soil and binder is quasi-statically compacted. Resin droplets are squeezed and driven by large capillary forces. Eventually, polymer micro-agglomerations (PMA) are self-assembled among adjacent filler grains. This nonuniform binder distribution maximizes the system efficiency of load transfer. With the binder content as low as 3~4 wt%, it leading to a high flexural strength around 20-40 MPa, stronger than most concrete parts with steel rebars. The compressive strength of the UBC composite is 50-60 MPa, on the same scale as high-strength concrete [26]. The polymer binder could be a thermoset, e.g. epoxy. The cost of epoxy is around $2-3 kg [27]. Processing of thermosets consumes energy and emits about 2-3 kg $CO_2$ per kg of product [28].

Due to the large size of the construction materials market, it is desirable to have an alternative binder source that does not increase the burden of polymer production. One attractive method is to use recycled waste plastics. In 2013, ~300 million tons of plastics are produced worldwide, which causes tremendous environmental concerns [29]. Common waste plastics include polyethylene, polystyrene, polypropylene, etc. Fabrication of construction parts is tolerate to the impurities and contaminations, circumventing the main reason why currently most of the plastics cannot be recycled [30]. Thermoplastics do not crosslink but would melt upon heating. In the current study, we investigate polyethylene (PE) as the binder. Once successful, it will offer a promising solutions not only to the mitigation of "white pollution" [31], but also to the reduction of energy use and carbon emission from the construction industry.

## 2. Experimental procedure

Polyethylene was obtained from Sigma-Aldrich (SKU 434272), with the particle size around 40-50 μm. All-purpose sand was provided by Quikrete (Product No. 1152), meeting the requirement of ASTM C33/33M. The largest sand grains (> 2 mm) were removed through sieve analysis. The sand was air dried at ambient temperature (~22 °C) for over 24 hours.

About 3 grams of the filler (sand) and the binder (PE) were weighed separately and transferred into a 40-ml beaker, and manually mixed by a steel spatula for 2 min (Fig.1A). The binder content ranged from 4% to 30%. The premixed material was poured into a cylindrical steel mold that had been preheated at 200 °C, and placed in a Jeio Tech OF-12G-120 convection oven at 200 °C. The heating time ($t_H$) ranged from 10 min to 1 h. The outer diameter, the inner diameter, and the height of the steel mold was 50.8 mm, 44.45 mm, and 19.05 mm, respectively. The two ends of the mold were end-capped by two preheated 25.4 mm-long 19.05 mm-diameter steel pistons.

The mold was moved out of the oven, and placed on the loading stage of an Instron type-5582 universal testing machine (Fig.1B). The loading rate was set to 15 mm/min. The upper piston was pushed into the mold, to reach the maximum compaction pressure ($P_c$). The value of $P_c$ ranged from 70 MPa to 350 MPa. The peak pressure was maintained for 1 min, after which the compacted material was pushed out of the mold and air cooled.

By using a high-speed diamond saw (MTI SYJ-40-LD), the produced PE-binder sample was sectioned into 18-mm-long 5-mm-large specimens. A set of 400-grit sand papers were used to polish the specimen surfaces. The flexural strength was measured in through three-point bending (Fig.1C). Two 20 mm-long 2 mm-diameter steel pins supported the specimen, with the span between the two pins ($L$) being ~16 mm. A compression pin was pressed downwards by the Instron machine, at the center of the upper surface. The compression rate was 6 mm/min. The flexural strength was $R = \frac{3}{2}\frac{F_f L}{bd^2}$, where $F_f$ is the peak compression force at specimen failure, $b$ is the specimen width, and $d$ is the specimen thickness. For each testing condition, at least 3 nominally same specimens were measured. Fractography was analyzed through a scanning electron microscope (SEM).

## 3. Result and discussions

Figure 2(A) shows the relationship of the measured flexural strength ($R$) and the PE content ($c$). A larger $c$ helps increase $R$, as it should. When the PE content is only 4 wt%, the flexural strength is ~8 MPa, nearly two times stronger than typical concrete [32]. With 7 wt% PE binder, $R$ is around 14 MPa, comparable to many steel-reinforced concrete [33]. The strength rapidly rises to ~23 MPa when $c$ is 10 wt%. As $c$ further increases, its beneficial effect becomes less pronounced. With $c = 25$ wt%, the optimum strength is achieved around 30 MPa, close to the inherent strength of PE [34], suggesting that the materials system has saturated. An even larger PE content would not further enhance $R$. However, when the binder content is above 30 wt%, the material is quite ductile. Clearly, as the PE phase fully occupy the free space among the filler grains, it dominates the overall mechanical properties.

Figure 2(B) demonstrates the effect of the peak compaction pressure ($P_c$). When $P_c$ is less than 140 MPa, increasing it is beneficial. As $P_c$ rises from 70 MPa to 140 MPa, $R$ is improved from ~22 MPa to ~27 MPa by nearly 22%. A larger compaction pressure above 140 MPa does not lead to a stronger material. In general, when the material is premixed (Fig.3A) and heated at 200 °C (Fig.3B), the sand grains cannot form bondings with each other, because the PE amount is small. The critical step is the CSA operation (Fig.3C). As the compaction loading is applied, the sand grains are pressed toward each other, associated with grain deformation, sliding, and rotation. As a result, the PE melt spreads over local areas. More critically, when the peak compaction pressure is maintained, a large capillary force presents at the narrowest microstructure sites. The capillary force pulls the PE melt to the contact places of adjacent sand grains, so that the binder is most efficiently utilized to bridge the filler together. It explains why the material is much stronger than regular composites of the same binder contents, particularly when $c$ is close to the lower end of the range of our investigation. With a higher compaction pressure, the short-range sand movement is promoted and the capillary force is larger, so that the binder dispersion is more uniform and the final strength is higher. This is demonstrated by the comparison between Fig.4(A) and Fig.4(B). When $c > 20$ wt% or $P_c > 140$ MPa, the PE distribution is quite homogeneous. As $R$ approaches the PE strength, no further improvement can be obtained, as shown by the comparison between Fig.4(B) and Fig.4(C). To maximize the production yield and to minimize the machinery complexity, the optimum PE content should be 7~10 wt% and the compaction pressure may be 70~100 MPa.

Since sand is not a good heat conductor, melting of PE powders takes time. As shown in Fig.5(A), as the heating time ($t_H$) is increased from 10 min to 30 min, the material strength is considerably improved from ~14 MPa to ~25 MPa by ~80%. We observed that as $t_H$ became longer, the PE color changed from white to somewhat grayish. If $t_H$ exceeded 30 min, the color turned slightly yellowish and correspondingly, the strength decreased. When $t_H$ was 50~60 min, $R$ was reduced back to ~14 MPa. It should be related to the oxidation and possible thermal decomposition of PE [34]. As the long chains are broken apart, the effective molecular weight is smaller, and the PE strength is detrimentally affected.

Another important processing parameter is the out time ($t_O$), i.e., the duration from the completion of heating to the onset of compaction. Its influence is shown in Fig.5(B). As the heating ends and the mold is moved out of the oven, the temperature begins to decrease. While no solidification takes place within a few minutes, the viscosity of the polymer melt rises [35]. The highest strength is achieved when the out time is less than 1 min. When $t_O$ becomes longer, $R$ decreases. As $t_O$ is 3~6 min, $R$ is only 40~60% of the peak strength. When $t_O$ is around 10 min, local solidification can be observed, and the strength largely decreases by ~80%, since the binder is poorly distributed and a relatively large number of defects are formed.

We also investigated the effect of the cooling rate. Instead of air cooling, after compaction, the material was immediately taken out of the mold and transferred to a temperature chamber at -5 ºC for 1 h, and then rested in ambient air for 12 h. The compaction pressure was 350 MPa; the heating time was 10 min; the out time was 1 min; the PE content was 4 wt%. The fast cooling does not have any evident influence on the flexural strength, indicating that the solidification process is not a critical procedure. The microstructure of the material is mainly formed during the CSA operation, before cooling starts.

In another set of tests, we confined the material during heating. As the mold was moved to the oven, the two pistons were firmly clamped, so that their positions were fixed. The rest of the procedure remained unchanged. The compaction pressure was 350 MPa; the heating time was 30 min; the out time was 1 min; and the PE content was 10 wt%. With the additional confinement, the material strength decreased from ~17 MPa to ~13 MPa by about 25%. This result suggests that free expansion is critical for the preparation of CSA. When PE melts, its volume considerably increases. If the system volume is fixed, the motion of the polymer melt is limited, and may be forced to the largest vacancies by the internal pressure. Hence, widespread dispersion of the binder phase is difficult.

## 4. Conclusions

In summary, we investigated thermoplastic-binder particular composites, with sand as the filler. Polyethylene (PE) was employed as the model material for waste plastics. The goal was to produce useful construction materials. After sand and PE powders were premixed, they were heated and then compacted. The compaction was not to form the material into shape, but to quasi-statically disperse the binder to the critical load-carrying microstructural sites. The material strength increases with the binder content and the compaction pressure. The optimum binder amount is ~10 wt%; the optimum compaction pressure is 70~100 MPa. The so-processed material has a flexural strength around 15 MPa, better than many steel-reinforced concrete. If the heating time is too short or too long, the strength would be reduced. The out time should be as short as possible. Cooling rate does not have an evident influence. Free expansion during heating is beneficial. Because construction materials have a low requirement on the binder purity, this study may enable large-scale recycling of waste plastics, and also help reduce the energy use and carbon emission from the construction industry.

**Acknowledgement**

This work was supported by ARPA-E under Grant No. DE-AR0001144. This work was performed in part at the San Diego Nanotechnology Infrastructure (SDNI) of the University of California – San Diego, a member of the National Nanotechnology Coordinated Infrastructure, which is supported by the National Science Foundation (Grant No. ECCS-1542148).

**Figures**

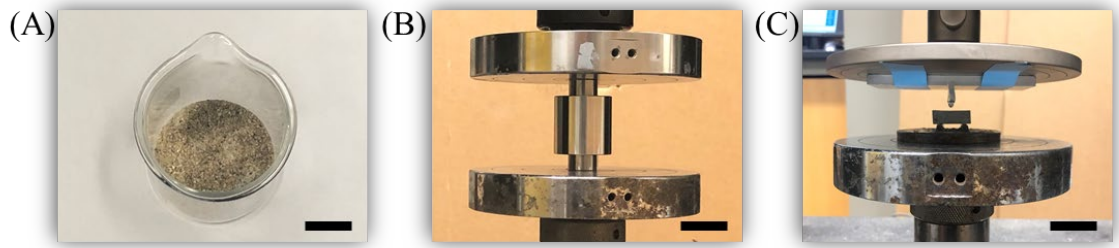

**Fig.1** Typical photos of (A) premixed filler (sand) grains and binder (PE) powders (scale bar: 10 mm), (B) mixed materials compacted in a steel mold (scale bar: 20 mm), and (C) three point bending test (scale bar: 20 mm).

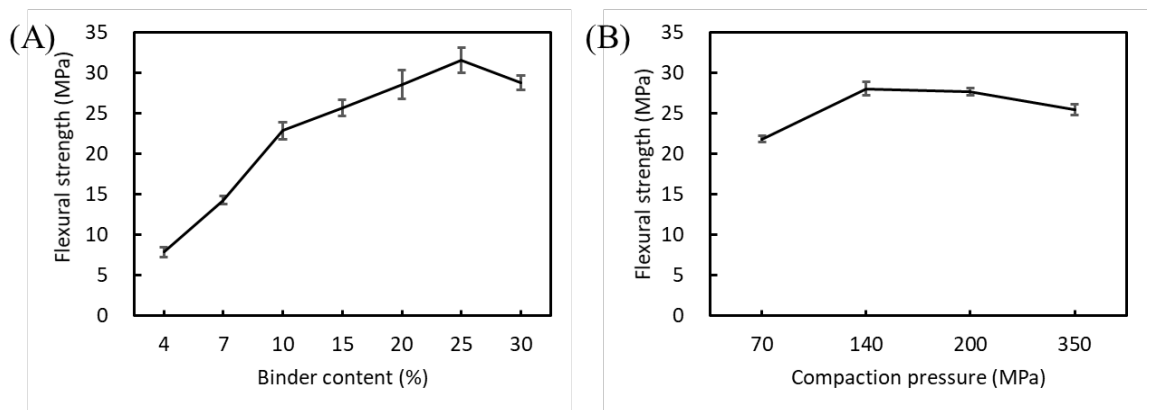

**Fig.2.** The flexural strength as a function of (A) the binder content and (B) the compaction pressure. The heating time is 30 min; the out time is 1 min; the compaction pressure in (A) is 350 MPa; the binder content in (B) is 10 wt%.

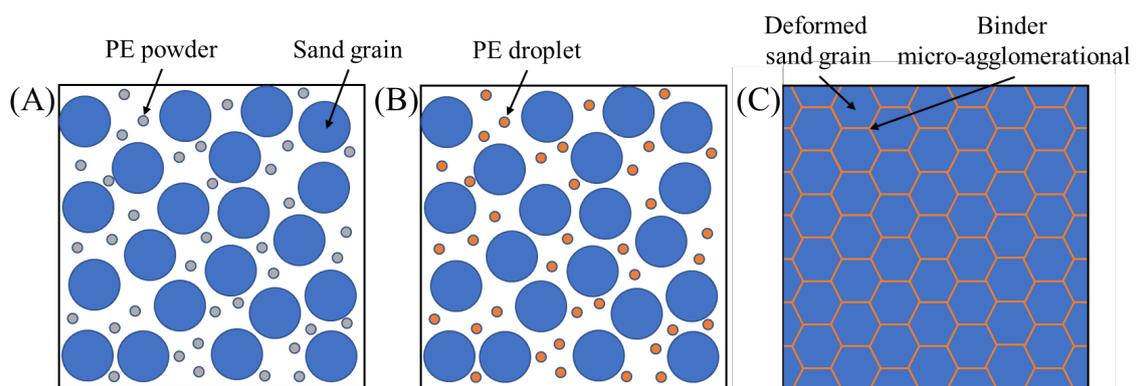

**Figure 3** Schematics showing (A) mixed sand grains (the filler) and PE powders (the binder); (B) upon heating, PE is melted; (C) upon compaction, the sand grains are deformed and densified, and the PE droplets are self-assembled into binder micro-agglomerations.

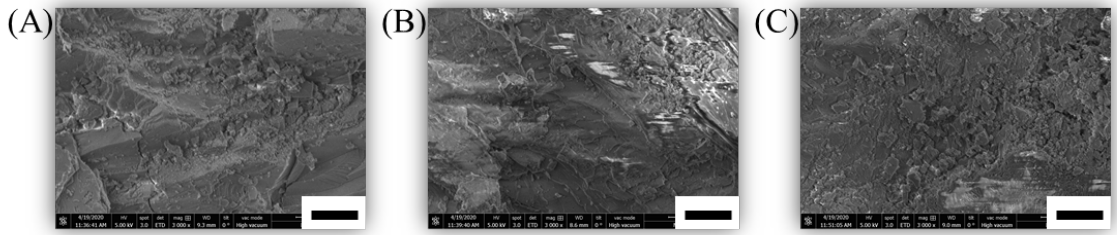

**Fig.4** SEM images of the PE-binder samples formed at the compaction pressures of (A) 70 MPa, (B) 140 MPa, and (C) 350 MPa (scale bar: 10 μm).

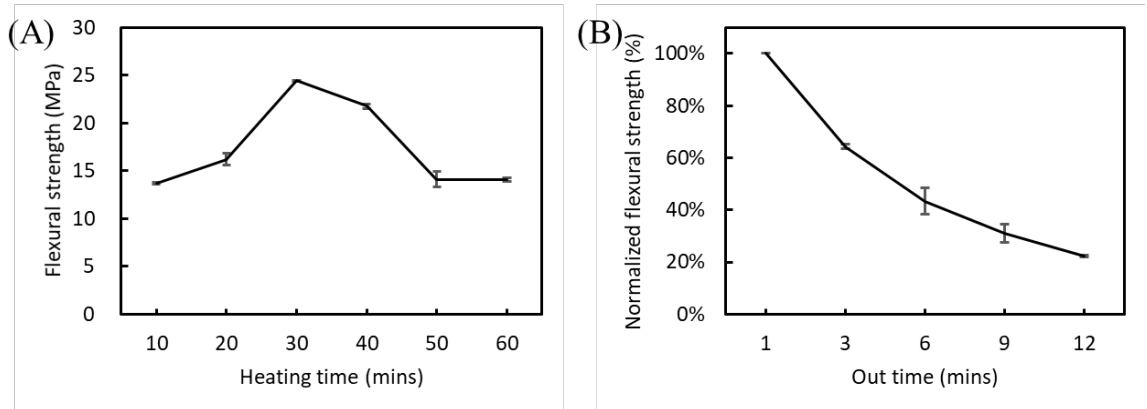

**Fig.5** The flexural strength as a function of (A) the heating time and (B) the out time. The compaction pressure is 350 MPa; the binder content is 10 wt%; the out time in (A) is 1 min; the heating time in (B) is 30 min.